\documentclass[conference,10pt]{IEEEtran}
\IEEEoverridecommandlockouts
\usepackage{amsmath,amssymb,amsfonts}
\usepackage[shortlabels]{enumitem}
\usepackage[lined,ruled,linesnumbered]{algorithm2e}
\SetKwComment{Comment}{/* }{ */}
\usepackage{algorithmic}
\usepackage{graphicx}
\usepackage{textcomp}
\usepackage{xcolor}
\usepackage{bm}
\usepackage{epsfig}
\usepackage{subfigure}
\usepackage{psfrag}
\usepackage{cite}
\def\BibTeX{{\rm B\kern-.05em{\sc i\kern-.025em b}\kern-.08em
		T\kern-.1667em\lower.7ex\hbox{E}\kern-.125emX}}

\begin{document}
\title{{\color{black} Convergence Analysis of Over-the-Air FL with Compression and Power Control via Clipping}}
\author{\IEEEauthorblockN{Haifeng Wen${}^{\ast}$, Hong Xing${}^{\ast\S}$, Osvaldo Simeone${}^\dagger$}\\
	\IEEEauthorblockA{${}^\ast$ The Hong Kong University of Science and Technology (Guangzhou), Guangzhou, China \\
		${}^\S$ The Hong Kong University of Science and Technology, HK SAR, China \\
        ${}^\dagger$ KCLIP Lab, CTR, Department of Engineering, King's College London, London, U.K.\\
		E-mails:~hwen904@connect.hkust-gz.edu.cn,~hongxing@ust.hk.,~osvaldo.simeone@kcl.ac.uk}
\thanks{The work of O. Simeone was supported by the European Research Council (ERC) through European Union’s Horizon 2020 Research and Innovation Programme under Grant 725731, by  an Open Fellowship of the EPSRC with reference EP/W024101/1, and by the European Union’s Horizon Europe project CENTRIC (101096379). }
}

\maketitle

\begin{abstract}

One of the key challenges towards the deployment of over-the-air federated learning (\emph{AirFL}) is the design of mechanisms that can comply with the power and bandwidth constraints of the shared channel, while causing minimum deterioration to the learning performance as compared to baseline noiseless implementations.  
For additive white Gaussian noise (AWGN) channels with instantaneous per-device power constraints, prior work has demonstrated the optimality of a power control mechanism based on norm clipping. This was done through the minimization of an upper bound on the optimality gap for {\color{black} smooth learning objectives satisfying the Polyak-\L{}ojasiewicz (PL) condition}. In this paper, we make two contributions to the development of AirFL based on norm clipping, which we refer to as \emph{AirFL-Clip}. First, we provide a convergence bound for AirFL-Clip that applies to general smooth and non-convex learning objectives. Unlike existing results, the derived bound is free from run-specific parameters, thus supporting an  \emph{offline} evaluation. Second, we extend AirFL-Clip to include Top-$k$ sparsification and linear compression. For this generalized protocol, referred to as \emph{AirFL-Clip-Comp}, we derive a convergence bound for general smooth and non-convex learning objectives. We argue, and demonstrate via experiments, that the only time-varying quantities present in the bound can be efficiently estimated offline by leveraging the well-studied properties of  sparse recovery algorithms. 


\end{abstract}


\IEEEpeerreviewmaketitle
\newtheorem{definition}{\underline{Definition}}[section]
\newtheorem{fact}{Fact}
\newtheorem{assumption}{Assumption}
\newtheorem{theorem}{\underline{Theorem}}[section]
\newtheorem{lemma}{\underline{Lemma}}[section]
\newtheorem{proposition}{\underline{Proposition}}[section]
\newtheorem{corollary}[proposition]{\underline{Corollary}}
\newtheorem{example}{\underline{Example}}[section]
\newtheorem{remark}{\underline{Remark}}[section]
\newcommand{\mv}[1]{\mbox{\boldmath{$ #1 $}}}
\newcommand{\mb}[1]{\mathbb{#1}}
\newcommand{\Myfrac}[2]{\ensuremath{#1\mathord{\left/\right.\kern-\nulldelimiterspace}#2}}
\newcommand\Perms[2]{\tensor[^{#2}]P{_{#1}}}

{\color{black}\section{Introduction}

As the number of edge devices increases, an enormous amount of private data is generated that cannot be directly used for training. To leverage this data while maintaining privacy, federated learning (FL) was proposed by Google in  \cite{mcmahan2017communication}. In FL, edge devices train a shared deep neural network (DNN) model using their local datasets, and the updated models are transmitted to a central server for global aggregation, which is then broadcast to the edge devices. Although FL provides solution for preserving certain level of data privacy, it encounters communication efficiency challenges. With DNN models being highly dimensional and the number of FL iterations typically being in hundreds, communication bottlenecks arise, particularly in wireless scenarios where communication resources are limited \cite{yang2020federated}.

To improve the efficiency of wireless FL, \emph{over-the-air computing (AirComp)} has been proposed to improve communication efficiency  for the global-model aggregation step of FL  \cite{yang2020federated, amiri2020federated, zhu20one-bit, wei2022federated}. The resulting protocols, referred to as  \emph{AirFL}  here, have been studied, for instance, in \cite{yang2020federated}, which proposed an optimization scheme for  device selection and beamforming to minimize the aggregation error. Digital-based and analog-based transmission AirFL schemes were proposed in \cite{amiri2020federated, xing2020decentralized}, where the authors introduced gradient/model  sparsification, along with compressed sensing, to address the mismatch between dimensions of gradients and available wireless resources. Aggregated signal detection methods based on compression sensing for AirFL were also presented in \cite{jeon2020compressive, jeon2022communication} in multi-antenna settings. Analyses of convergence for convex loss functions can be found in \cite{sery2021over} and \cite{guo2020analog}. Bayesian learning via AirComp was proposed in \cite{liu2022wireless}, while privacy aspects were studied in \cite{liu20privacy}.

One of the key challenges towards the deployment of AirFL is the design of mechanisms that can comply with the power and bandwidth constraints of the shared channel, while causing minimum deterioration to the learning performance as compared to baseline noiseless implementations.  
For additive white Gaussian noise (AWGN) channels with instantaneous per-device power constraints, prior work has demonstrated the optimality of a power control mechanism based on norm clipping, which minimized an upper bound on the optimality gap for learning objectives satisfying Polyak-\L{}ojasiewicz (PL) condition and Lipschitz smoothness \cite[Theorem 1]{cao2021optimized}. 


In this paper, we make two contributions to the development of AirFL based on norm clipping, which we refer to as \emph{AirFL-Clip}. \begin{itemize} \item First, we provide a convergence bound for AirFL-Clip that applies to general smooth and non-convex learning  objectives. Unlike existing results \cite{cao2021optimized, xing2021federated}, the derived bound  is free from run-specific parameters, thus supporting an  \emph{offline} evaluation. \item Second, we extend AirFL-Clip to include  Top-$k$ sparsification and linear compression. For this generalized protocol, referred to as \emph{AirFL-Clip-Comp}, we present a convergence bound for general smooth and non-convex learning objectives. We argue, and demonstrate via experiments, that the only time-varying quantities present in the bound can be efficiently estimated offline by leveraging the well-studied tools for  sparse recovery algorithms, such as state evolution (SE) \cite{donoho2009message, ma2017orthogonal}. 
\end{itemize}

The rest of the paper is organized as follows. Sec. II presents the system model, and Sec. III introduces AirFL-Clip, along with its extension AirFL-Clip-Comp that encompasses compression.  Sec. IV describes the main analytical results; Sec. V puts forth experimental results; and Sec. VI concludes this paper and discusses future work.

}

\section{System Model} \label{sec:System Model} 

\begin{figure}[tp]
    \centering
    \includegraphics[width=3.2in]{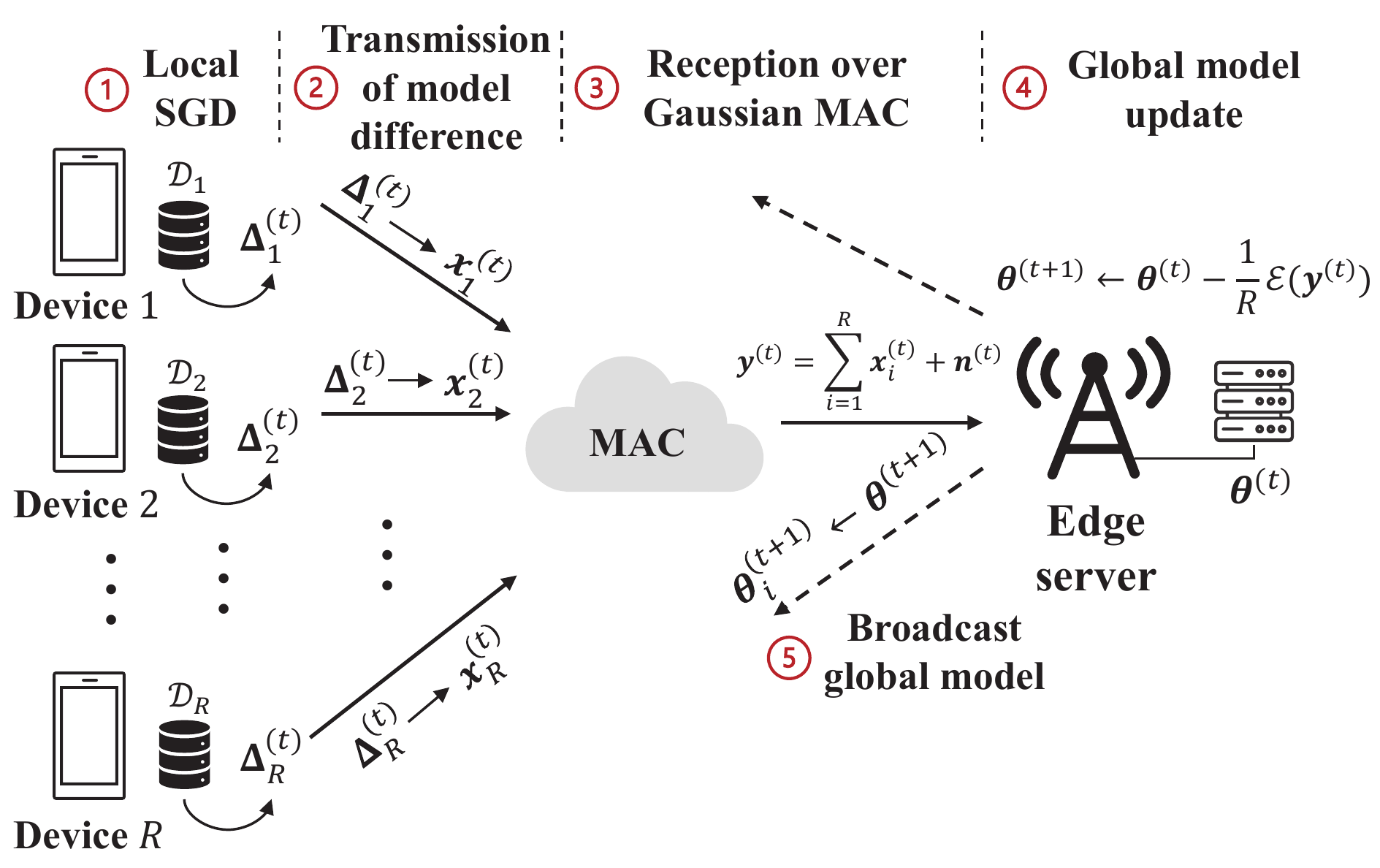}
    \caption{Illustration of an AirFL system.}\label{fig:system model}
    \vspace{-.2in}
\end{figure}

As shown in Fig. \ref{fig:system model}, we consider a wireless FL system consisting of an edge server and $R$ devices, all equipped with single antennas.  A common machine learning model is collaboratively trained by all devices through analog communications with the edge server over a shared noisy channel. We focus on studying the impact of a novel power control strategy based on norm clipping. In this section, we introduce the learning protocol and the communication model.

\subsection{Learning Protocol (Vanilla FL)} \label{subsec:Learning Model}
Each device $i$ in the considered FL system possesses a local dataset denoted as $\mathcal{D}_i$ for $i\in [R]=\{1,...,R\}$. All devices share a common machine learning model, such as a DNN, which is parameterized by a vector $\mv \theta \in \mb{R}^{d \times 1}$. The objective of this network is to collaboratively solve the empirical loss minimization problem $\mathrm{(P0)}$, which is given by
\begin{align*}
\mathrm{(P0)}:&~\mathop{\mathtt{Minimize}}_{\mv \theta}~~~f(\mv \theta)\triangleq\frac{1}{R}\sum_{i=1}^{R}f_i(\mv \theta),
\end{align*} 
where $f(\mv \theta)$ represents the global empirical loss function; $f_i(\mv \theta) = 1/\vert\mathcal{D}_i\vert\sum_{\bm\xi\in\mathcal{D}_i}\mathcal{L}(\mv \theta;\bm\xi)$ is the local empirical loss function for device $i\in[R]$ with \(|\cdot|\) denoting the cardinality of a given set; and $\mathcal{L}(\mv \theta;\bm\xi)$ denotes the loss function for the parameter $\mv \theta$ evaluated on the data sample $\bm \xi$. 

Next we describe the \emph{vanilla FL} scheme that assumes ideal, \emph{noiseless} communications between multiple devices and a server \cite{mcmahan2017communication}. Denote the index of a global communication round by \(t\in\{0,\ldots, T-1\}\). At the $t$-th \emph{global iteration}, each device has a local parameter vector $\mv \theta_i^{(t)}$ that approximates the solution \(\mv \theta^{(t)}\) to problem (P0). At each global iteration $t$, each device performs $Q$ local stochastic gradient descent (SGD) steps on its dataset $\mathcal{D}_i$, thereby updating its local model parameter $\mv \theta_i^{(t)}$ as
\begin{align}
\mv \theta_i^{(t, q+1)} \leftarrow \mv \theta_i^{(t, q)}-\eta^{(t)}\hat\nabla f_{i}(\mv \theta_i^{(t, q)}), \label{eq:local updates}
\end{align} 
where $q\in\{0,\ldots, Q-1\}$ is the \emph{local iteration} index; $\eta^{(t)}$ denotes the learning rate; and $\hat\nabla f_{i}(\mv \theta_i^{(t,q)})$ is the estimate of the true gradient $\nabla f_i(\mv \theta_i^{(t,q)})$ obtained from a mini-batch $\mathcal{D}_i^{(t)}\subseteq\mathcal{D}_i$ of device $i$'s dataset as 
\begin{align}
\hat\nabla f_{i}(\mv \theta_i^{(t, q)})=\frac{1}{\vert\mathcal{D}_i^{(t)}\vert}\sum\limits_{\bm\xi\in\mathcal{D}_i^{(t)}}\nabla \mathcal{L}(\mv \theta_i^{(t, q)};\bm\xi). \label{eq:SGD}
\end{align} The local training iterations (\ref{eq:SGD}) are initialized with the common parameter vector \(\mv \theta_i^{(t,0)}=\mv \theta_i^{(t)}\leftarrow \mv \theta^{(t)}\) shared by the edge server, as described next.

The edge server receives the \emph{model differences} \begin{equation}\label{eq:modeldiff}\mv{\Delta}_i^{(t)} = \mv \theta_i^{(t,0)}-\mv \theta_i^{(t, Q)}\end{equation} from the $R$ devices, and it aggregates them to update the global model parameters as \begin{equation}
    \mv \theta^{(t+1)} \leftarrow \mv \theta^{(t)} - \frac{1}{R}\sum_{i=1}^{R}\mv{\Delta}_i^{(t)}. \label{eq:global updates}
\end{equation} This updated model parameter is then broadcast to all $R$ devices, who use it to initialize the local iterates \eqref{eq:local updates}, i.e.,  $\mv \theta_i^{(t+1)} \leftarrow \mv \theta^{(t+1)} $, \(i\in[R]\). The steps \eqref{eq:local updates} and \eqref{eq:global updates} are iterated until certain convergence criteria are met.

\subsection{Communication Model} \label{subsec:Communication Model}
Unlike the vanilla FL protocol reviewed in the previous section, in this paper we study a wireless system in which each device \(i\in[R]\) transmits its model difference $\mv{\Delta}_i^{(t)}$ in (\ref{eq:modeldiff}) to the edge server over a Gaussian multiple access channel (MAC) via {over-the-air computing (AirComp)} \cite{yang2020federated,amiri2020federated}.

To this end, first, the $d$-dimension model difference $\mv{\Delta}_i^{(t)}$ is mapped into an $M$-dimension vector ${\mv{x}}_i^{(t)}$, where $M$ represents the number of channel uses available for transmission at each global communication round. If $M<d$, compression is applied by means of sparsification and linear projection operations as elaborated in the next Section.

All $R$ devices transmit simultaneously over a Gaussian MAC, and the edge server receives an $M\times 1$ vector given by
\begin{equation} \label{eq:rxsignal}
    \mv{y}^{(t)}=\sum\limits_{i=1}^{R} {\mv{x}}_i^{(t)}+\mv{n}^{(t)},
\end{equation}
where the instantaneous transmit power $||{\mv{x}}_i^{(t)}||^2$ per communication block for device $i\in[R]$ must satisfy the constraint $||{\mv{x}}_i^{(t)}||^2\leq P_{i, \max}$, and $\mv{n}^{(t)}\sim \mathcal{N}(\mv 0, \sigma^2 \mv I)$ is the additive white Gaussian noise (AWGN) channel-noise vector. Based on the received signal (\ref{eq:rxsignal}), the edge server estimates  the update term in \eqref{eq:global updates}, namely \(\sum_{i=1}^R\mv\Delta_i^{(t)}\), to obtain the global model parameter vector $\mv{\theta}^{(t+1)}$. As in \cite{amiri2020federated}, we assume ideal and noiseless communication in the downlink, given the less constrained resources available for downlink communication at the edge server.

\section{AirFL with Power Control via Clipping} \label{sec:FedAcc}
{  
In this section, we describe AirFL-Clip \cite{cao2021optimized}, as well as its extension AirFL-Clip-Comp, which incorporates sparsification and linear projection.

\subsection{AirFL-Clip}
Many papers on AirFL, such as \cite{yang2020federated, amiri2020federated, jeon2020compressive, jeon2022communication, guo2020analog, xing2021federated}, limit the transmitted power by scaling the model differences -- or compressed versions thereof -- as $\mv x^{(t)}_i=\sqrt{\alpha_i}\mv{\Delta}^{(t)}_i$ for some coefficients $\alpha_i>0$, {\color{black} ensuring that all signals are transmitted subject to the given power budget $P=\min_{i\in [R]}\{P_{i,\max}\}$} for signal alignment at the edge server}. In contrast, AirFL-Clip \cite{cao2021optimized} limits the power of the transmitted signal by \emph{norm clipping} as 
\begin{equation}\label{eq:clipping1}
    \mv{x}_{i}^{(t)}=\text{clip}\left(\frac{1}{\eta^{(t)}}\mv{\Delta}_{i}^{(t)},\sqrt{P}\right),
\end{equation}
with the clipping function defined as\footnote{We use $\|\cdot\|$ to denote the $2$-norm in vector space unless otherwise specified.} 
\begin{equation}\label{eq:clipping function}
    \text{clip}(\mv{x},\sqrt{P})=\min\left\{1,\frac{\sqrt{P}}{\|\mv{x}\|}\right\}\mv{x}.
\end{equation} 


\subsection{AirFL-Clip-Comp} \label{sec:AirFL-Clip-Comp}
When the number of channel uses $M$ is smaller than the model dimension $d$, i.e., when $M<d$, it is necessary to compress the model differences $\mv{\Delta}^{(t)}_i$ prior to scaling or clipping for analog transmission. In this subsection we describe a direct extension of AirFL-Clip that incorporates sparsification and linear compression as done in, e.g., \cite{amiri2020federated}, for scaling-based protocols. 

At global iteration $t$, each device first sparsifies the model difference $\mv \Delta_i^{(t)}$ via  a \emph{Top-$k$ contraction} as \cite{basu2020qsparse}
\begin{equation} \label{eq:Top-k}
\mv{g}^{(t)}_i=\text{Top}_k(\mv{m}_{i}^{(t)}+\mv{\Delta}_{i}^{(t)}),
\end{equation}
which preserves only the entries with $k$ largest absolute values while setting others to zero. In \eqref{eq:Top-k}, a  $d\times 1$  memory vector $\mv{m}_{i}^{(t)}$ is used to keep track of the accumulated errors as 
\begin{equation}
    \mv{m}_{i}^{(t+1)}= \mv{m}_{i}^{(t)}+\mv{\Delta}_{i}^{(t)} - \mv g_i^{(t)}.
\end{equation}
Using such memory vector can intuitively retain the information lost at the current communication round for future iterations. The sparsified model difference $\mv g_i^{(t)}$ is then clipped as in \eqref{eq:clipping1}, producing   \begin{equation} \tilde{\mv x}_i^{(t)} = \operatorname{clip}\left( \frac{1}{\eta^{(t)}}\mv g_i^{(t)}, \sqrt{P} \right).\end{equation}


As in \cite{amiri2020federated}, each device projects the clipped vector $\tilde{\mv x}_i^{(t)}$ using the same  $M\times d$ matrix $\mv A^{(t)}$ to obtain the transmitted signal $ \mv x_i^{(t)} = \mv A^{(t)} \tilde{\mv{x}}_i^{(t)}$. Matrix $\mv A^{(t)}$ has the property that its spectral norm $\|\mv A^{(t)}\|_2$, i.e., the  square root of the maximum eigenvalue of matrix $\mv A^{(t)}(\mv A^{(t)})^T$, satisfies  the inequality $\|\mv A^{(t)}\|_2 \le 1$. This condition guarantees the power constraint  $\|{\mv{x}}_i^{(t)}\|= \|\mv A^{(t)} \tilde{\mv x}_i^{(t)}\| \le \| \mv A^{(t)}\|_2 \|\tilde{\mv x}_i^{(t)}\| \le \|\mv x_i^{(t)}\|\le \sqrt{P}$. Such a compression matrix  can be generated by, e.g., selecting $M$ rows of any unitary matrix. 

Like \eqref{eq:rxsignal}, the vector received at the server can be written as
\begin{equation} \label{eq:received signal}
    \mv{y}^{(t)}=\mv{A}^{(t)}\tilde{\mv{x}}^{(t)}+\mv{n}^{(t)},
\end{equation}
where $\tilde{\mv{x}}^{(t)} = \sum_{i=1}^{R} \tilde{\mv{x}}_i^{(t)}$ is the sum of the sparsified and clipped model differences. This signal is used to produce an estimate  $\hat{\mv x}^{(t)} = \mathcal{E}(\mv{y}^{(t)})$ of the sum $\tilde{\mv x}^{(t)}$. With such an estimate, the model parameter vector is finally updated as
\begin{equation} \label{eq:clip global updatas}
    \mv \theta^{(t+1)} \leftarrow \mv \theta^{(t)} - \eta^{(t)}\frac{1}{R}\hat{\mv x}^{(t)}.
\end{equation}
The AirFL-Clip-Comp scheme is summarized in Algorithm \ref{alg:Algorithm 1}. When $k=d$ and $\mv A = \mv I_d$, it reduces to AirFL-Clip.

\begin{algorithm}[t] \label{alg:Algorithm 1}
\SetKwInOut{Input}{Input}
\SetKwInOut{Output}{Output}
\SetKwBlock{DeviceParallel}{On devices $i \in [R]$:}{end}
\SetKwBlock{localSGD}{for $q=0$ to $Q-1$ do}{end}
\SetKwBlock{OnServer}{On Server:}{end}
\caption{AirFL-Clip-Comp}\label{alg:Algorhtm}
\Input{ $\{\eta^{(t)}\}_{t=0}^{T-1},~\sqrt{P},~k,~T,~Q$ and $\mathcal{E}(\cdot).$ }
Initialize $\mv \theta_i^{(0)}=\mv \theta^{(0)},~\mv m_i^{(0)} = \mv 0, ~\text{for all}~i \in [R],~t=0 $. \\
\While{$t < T$}{
\DeviceParallel{
$\mv{\theta}^{(t)}_{i}\leftarrow \mv{\theta}^{(t)}$\;
\localSGD{
$\mv{\theta}^{(t,q+1)}_{i}\leftarrow \mv{\theta}^{(t,q)}_{i} - \eta^{(t)} \hat\nabla f(\mv \theta_{i}^{(t,q)})$
}
$\mv{g}^{(t)}_i\leftarrow \text{Top}_k(\mv{m}_{i}^{(t)}+\mv \theta_i^{(t)}-\mv \theta_{i}^{(t,Q-1)};~k)$\;
$\mv{m}_{i}^{(t+1)}=\mv{m}_{i}^{(t)}+\mv \theta_i^{(t)}-\mv \theta_{i}^{(t,Q-1)}-\mv{g}_{i}^{(t)}$\;
$\tilde{\mv{x}}_{i}^{(t)}=\text{clip}\left(\frac{1}{\eta^{(t)}}\mv{g}_{i}^{(t)},\sqrt{P}\right)$\;
Obtain $\mv x_i^{(t)} =\mv A^{(t)} \tilde{\mv x}_i^{(t)}$ and transmit it to the server\;
}
\OnServer{
Receive $\mv{y}^{(t)}=\mv{A}^{(t)}\tilde{\mv{x}}^{(t)}+\mv{n}^{(t)}$\;
Sparse Recovery: $\hat{\mv x}^{(t)} = \mathcal{E}(\mv{y}^{(t)})$\; 
Global Update: $\mv \theta^{(t+1)}\leftarrow \mv \theta^{(t)}-\eta^{(t)} \frac{1}{R} \hat{\mv x}^{(t)} $\;
Broadcast $\mv \theta^{(t+1)}$ to $R$ devices\;
}
$t \leftarrow t + 1$\;
}
\end{algorithm}

\section{Main Results} \label{sec:Main Results}
In this section, we analyze the convergence of AirFL-Clip-Comp, also including the analysis of AirFL-Clip as a special case, for general smooth and non-convex loss functions. For AirFL-Clip, unlike prior art, the convergence result is free from realization-specific parameters, and can hence be evaluated offline; while for AirFL-Clip-Comp we demonstrate that the derived bound can be evaluated by leveraging the properties of sparse reconstruction algorithms.
\subsection{Convergence Analysis} \label{subsec:convergence results}
Our analysis is based on the following assumptions.

\begin{assumption}[$L$-smoothness] \label{assumption:L-smoothness}
    The local empirical loss function \(f_i(\mv\theta)\), \(i\in[R]\), is continuously differentiable, and its gradient \(\nabla f_i(\mv\theta)\) is Lipschitz such that for all pairs $ \mv \theta, \mv\theta^\prime \in \mb{R}^{d\times 1}$, there exists a constant $L$ satisfying
    \begin{equation} \label{eq:L-smoothness}
        f_i(\mv\theta^\prime)\le f_i(\mv\theta) + \left< \nabla f_i(\mv\theta), \mv\theta^\prime - \mv\theta\right> + \frac{L}{2}\|\mv\theta^\prime - \mv\theta\|^2,
    \end{equation}where $\left< \mv a, \mv b\right> = \mv a^T \mv b$ is the inner product.
\end{assumption}

\begin{assumption}[Bounded Gradient Variance and Norm] \label{assumption:Bounded Instantaneous Norm}
    The mini-batch local gradient \(\hat{\nabla} f_i(\mv\theta)\), \(i\in[R]\), is unbiased, i.e., \(\mb {E}[\hat\nabla f_{i}(\mv\theta)]=\nabla f_i(\mv\theta)\), with its variance bounded as
    \begin{equation}
        \mb {E}\left\| \hat\nabla f_{i}(\mv\theta) - \nabla f_i(\mv\theta) \right\|^2 \le \sigma_l^2,
    \end{equation}
    and 
    \begin{equation}
        \| \hat{\nabla} f_i(\mv\theta) \| \le G,
    \end{equation}where the average is taken with respect to the choice of the mini-batch.
\end{assumption}

\begin{assumption}[Bounded Data Heterogeneity] \label{assumption:Bounded Data Heterogeneity}
The local gradient \({\nabla} f_i(\mv\theta)\), \(i\in[R]\), satisfies
    \begin{equation}
        \|\nabla f_i(\mv\theta)-\nabla f(\mv\theta)\|^2 \le \sigma_g^2.
    \end{equation}
\end{assumption}

\begin{assumption}[Estimation Error] \label{assumption:Estimation Error}
The estimate $\hat{\mv x}^{(t)}=\mathcal{E}(\mv y^{(t)})$ can be expressed as  $\hat{\mv x}^{(t)} = \tilde{\mv x}^{(t)} + \mv n_{e}^{(t)}$, where the estimation error $\mv n_{e}^{(t)}$ has zero mean and variance $\mathbb{E}||\mv n_{e}^{(t)}||^2=d \cdot v^{(t)}$, for some $v^{(t)}>0$, and is uncorrelated with the signal $\tilde{\mv x}^{(t)}$. 
\end{assumption}

The last assumption is trivially satisfied under the AirFL-Clip scheme by setting the estimation error $v^{(t)}$ to be equal to the channel noise power, i.e., $ v^{(t)}= \sigma^2$ for all $t$. In the more general case of  AirFL-Clip-Comp, this assumption is satisfied by {\color{black}applying linear minimum mean squared error (LMMSE) estimators}. Furthermore, it is also, approximately, met by Bayesian sparse recovery algorithms \cite{ ma2017orthogonal}. Furthermore, for such Bayesian estimators, the sequence ${v^{(t)}}$ is a function only of the compression matrix $\mv A^{(t)}$, of the channel noise power $\sigma^2$, and of the  distribution of the signal $\mv{x}^{(t)}$ \cite{ma2017orthogonal}.  For instance, for approximate message passing (AMP) based schemes, one can evaluate the estimation error $v^{(t)}$ by leveraging SE  \cite{donoho2009message, ma2017orthogonal}.  {\color{black} The distribution of the signal $\mv{x}^{(t)}$  can be practically approximated by sparse distributions, e.g., Bernoulli Gaussian} (see Sec. V).

\begin{theorem}[Convergence] \label{thm:convergence}
Under  Assumptions 1-4, let $\{\mv{\theta}^{(t)} \}_{t=0}^{T-1}$ be generated according to Algorithm \ref{alg:Algorithm 1} with given parameters $\lambda=k/d$, $\sqrt{P}$, and $Q$. Then, given decaying learning rate $\eta^{(t)}=\xi/(a+t)$, where $a\ge \max\{4Q/\lambda, \sqrt{120}\xi Q L\}$ and a constant $C\ge \Myfrac{4a\lambda(1-\lambda^2)}{(a\lambda-4Q)}$, on average over the randomness of  SGD and of the channel noise,  AirFL-Clip-Comp satisfies the inequality \eqref{eq:bound} (seen at the top of the next page),
\begin{figure*}[t]
    \begin{equation} \label{eq:bound}
    \begin{aligned}
    \frac{1}{P_T} \sum_{t=0}^{T-1}\eta^{(t)}  \mathbb{E}\left[\left\|\nabla f(\mv\theta^{(t)})\right\|^{2}\right] 
    & \le \underbrace{\frac{4}{\Gamma Q P_T}\left(f(\mv\theta^{(0)})- f^* \right)}_{\text{Initialization error}} + \underbrace{\frac{20L^2Q(\sigma_l^2+6Q\sigma_g^2)}{\Gamma P_T}\frac{\xi^3}{2(a-1)^2}}_{\text{Local SGD \& non-i.i.d. error}}  + \underbrace{\frac{2dL}{\Gamma QR^2P_T}\sum_{t=0}^{T-1}(\eta^{(t)})^2 v^{(t)}}_{\text{Sparse recovery error}}\\
    & +\underbrace{\frac{64CG^2}{\Gamma \lambda^2} + 18L \left( \frac{8C}{\lambda^2}+2\right) Q G^2\frac{1}{\Gamma P_T}\frac{\xi^2}{a-1} +\frac{8G^2}{\Gamma }\sqrt{ \frac{8C}{\lambda^2}+2}}_{\text{Sparsification and clipping error}}.  \\
    \end{aligned}
    \end{equation}
    \hrulefill
    \vspace{-0.2in}
\end{figure*}
where $P_T=\sum_{t=0}^{T-1}\eta^{(t)} \ge \xi \ln((T+a-1)/a)$ and $\Gamma= \sqrt{P}/(\sqrt{P}+\sqrt{\frac{8C}{\lambda^2}+2}QG)$. 
\end{theorem}


\begin{IEEEproof}
A sketch of the proof is provided in the appendix.
\end{IEEEproof}

The bound in (\ref{eq:bound}) is annotated with different impairments that cause each term in the sum, including initialization, drift due to local SGD steps and non-i.i.d. data, sparsification and clipping error, and sparse recovery error. The special case of AirFL-Clip is obtained by setting $\lambda = 1$,  resulting in $C=0$, and, as discussed earlier in this section,  $v^{(t)}=\sigma^2$. The resulting  bound  (\ref{eq:bound})  does not depend on time-varying quantities, and thus can be evaluated offline, unlike the results in \cite[Theorem 1]{cao2021optimized} and \cite[Lemma 1]{guo2020analog}. 

More generally, for AirFL-Clip-Comp, the only term that depends on time-varying quantities is the last one involving the sum \(\sum_{t=0}^{T-1}(\eta^{(t)})^2v^{(t)}\), which is determined by the sequence of estimation errors. By Assumption 4, this term can be efficiently estimated, and we will show the merit of this bound via a numerical example in the next section. 

{\color{black} As a further connection with existing art, we observe that, with noiseless communications, i.e., with zero  estimation error (\(v^{(t)}=0\)), we obtain the known convergence rate $\mathcal{O}\left(1/\ln T\right)$ \cite[Theorem 2]{basu2020qsparse}, plus error-floor term $\frac{64CG^2}{\Gamma\lambda^2}+\frac{8G^2}{\Gamma}\sqrt{ \frac{8C}{\lambda^2}+2}$, which is induced by the joint impact of Top-$k$ contraction and of norm-clipping operation. 



\addtolength{\topmargin}{0.07in}
{\color{black}\section{Experiments and Discussions}\label{sec:Experiments}
In this section, we evaluate the performance of  AirFL-Clip-Comp  in a wireless setup with $R=20$ devices, with the main aim of comparing the performance with scaling-based schemes and validating the relevance of the proposed bound. 


We consider the learning task of classifying $10$ classes of fashion articles over the standard Fashion-MNIST dataset, which is divided into $60,000$ training data samples ($6,000$ samples per class) and $10,000$ test data samples with each of them corresponding to a $28\times28$ image. Each device has training data for all classes excluding $4$ classes (non-i.i.d.), with an equal number of samples for each available class. The samples are drawn randomly ({\color{black}with replacement}) from the training set. All devices train a common DNN model that consists of one input layer with input shape $(28, 28)$, two fully-connect layers of $32$ and $16$ nodes, respectively, with ReLU activation function followed by a dropout layer with $\text{rate}=0.2$, and a softmax output layer, yielding a total number $d= 25,818$ of training parameters. 

The simulation parameters are set as follows unless otherwise specified: the mini-batch size is $\vert\mathcal{D}_i^{(t)}\vert=128$; the number of local iterations $Q = 1$; the power constraint \(P = P_{i,\max}=2\times 10^{-5}d\) for all $i\in [R]$; the SNR $P/(d\sigma^2)=30$\,dB;  and the sparsity level for Top-$k$ operator $k/d=0.1$ for each device. {\color{black} We adopt the orthogonal AMP (OAMP) \cite{ma2017orthogonal} as the estimator $\mathcal{E}(\cdot)$ (see Assumption 4). When deriving the estimator to evaluate the derived bound in Theorem 4.1,  we set the distribution of the signal $\tilde{\mv x}^{(t)}$ as i.i.d. Bernoulli-Gaussian  with sparsity level  $0.1$. Accordingly, for each entry of vector $\tilde{\mv x}^{(t)}$, we assume a distribution $p(x)=0.9\delta(x)+0.1\mathcal{N}(0, P/d)$, where $\delta(\cdot)$ is the Dirac measure at $0$. The number of iterations of OAMP is set as $20$.}

As benchmarks, we consider the following schemes.
\begin{enumerate}
   \item AirFL-Scale: Clipping is replaced with power scaling with the scaling factor $P_{i,\max}/\|\Delta_i^{(t)}\|^2$, while all other operations remain the same as in AirFL-Clip \cite{cao2021optimized}.
    \item AirFL-Scale-Comp: This scheme follows AirFL-Scale while including also compression via Top-$k$ and linear projection (see Sec. \ref{sec:AirFL-Clip-Comp}).
    \item Vanilla FL: This scheme refers to the standard FL algorithm \cite{mcmahan2017communication}.
\end{enumerate}


We first compare the training loss of the mentioned schemes versus the communication round $T$, as shown in Fig. \ref{fig:loss_vs_T}. Note that the performance of Vanilla FL with perfect communication provides a lower bound on the performance of protocols operating over noisy channels. The main observation is that clipping schemes generally achieve better performance than scaling-based benchmarks. 


\begin{figure}[htp]
    \centering
    \includegraphics[width=3.2in]{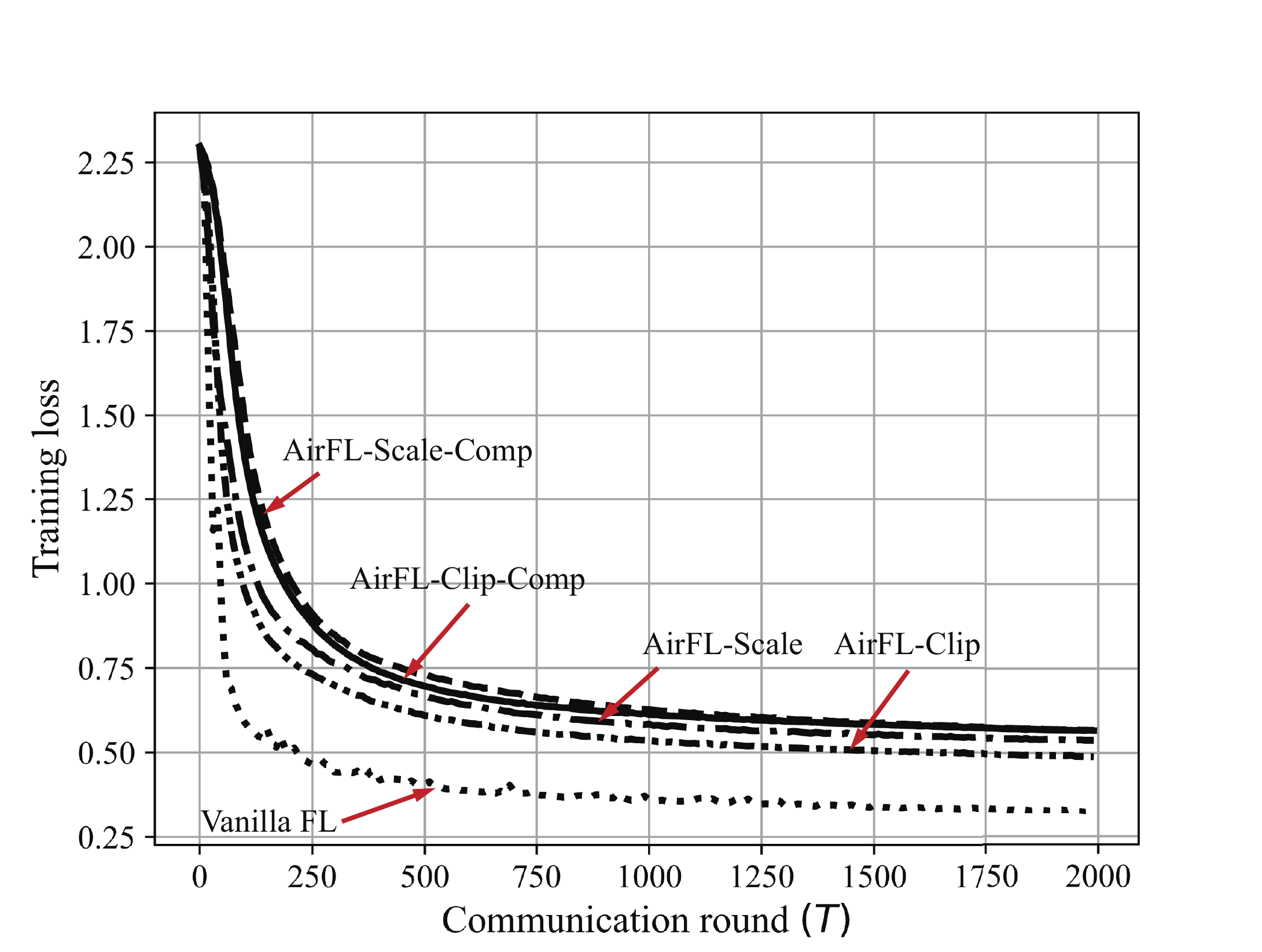}
    \vspace{-0.1in}
    \caption{Training loss performance of AirFL-Clip-Comp, AirFL-Scale with compression, AirFL-Scale, and Vanilla FL versus $T$ with $\eta^{(t)}=120/(300+t)$ and $M/d=0.6$ for AirFL-Clip and AirFL-Scale. }
    \label{fig:loss_vs_T}
    \vspace{-0.1in}
\end{figure}

Furthermore, to elaborate on the relevance of the derived analytical bound, Fig. \ref{fig:loss_vs_M_d} shows how the empirical loss varies as a function of the dimension-reduced level $M/d$ for AirFL-Clip-Comp, along with the theoretical bound \eqref{eq:bound}. Note that we have rescaled the bound by its corresponding value at $t =0$ and offset the constant terms for the ease of comparison.
It is observed that the bound predicts  well the minimum number of communication resources required to achieve a prescribed value of the loss function.


\begin{figure}[htp] 
    \centering
    \includegraphics[width=3.2in]{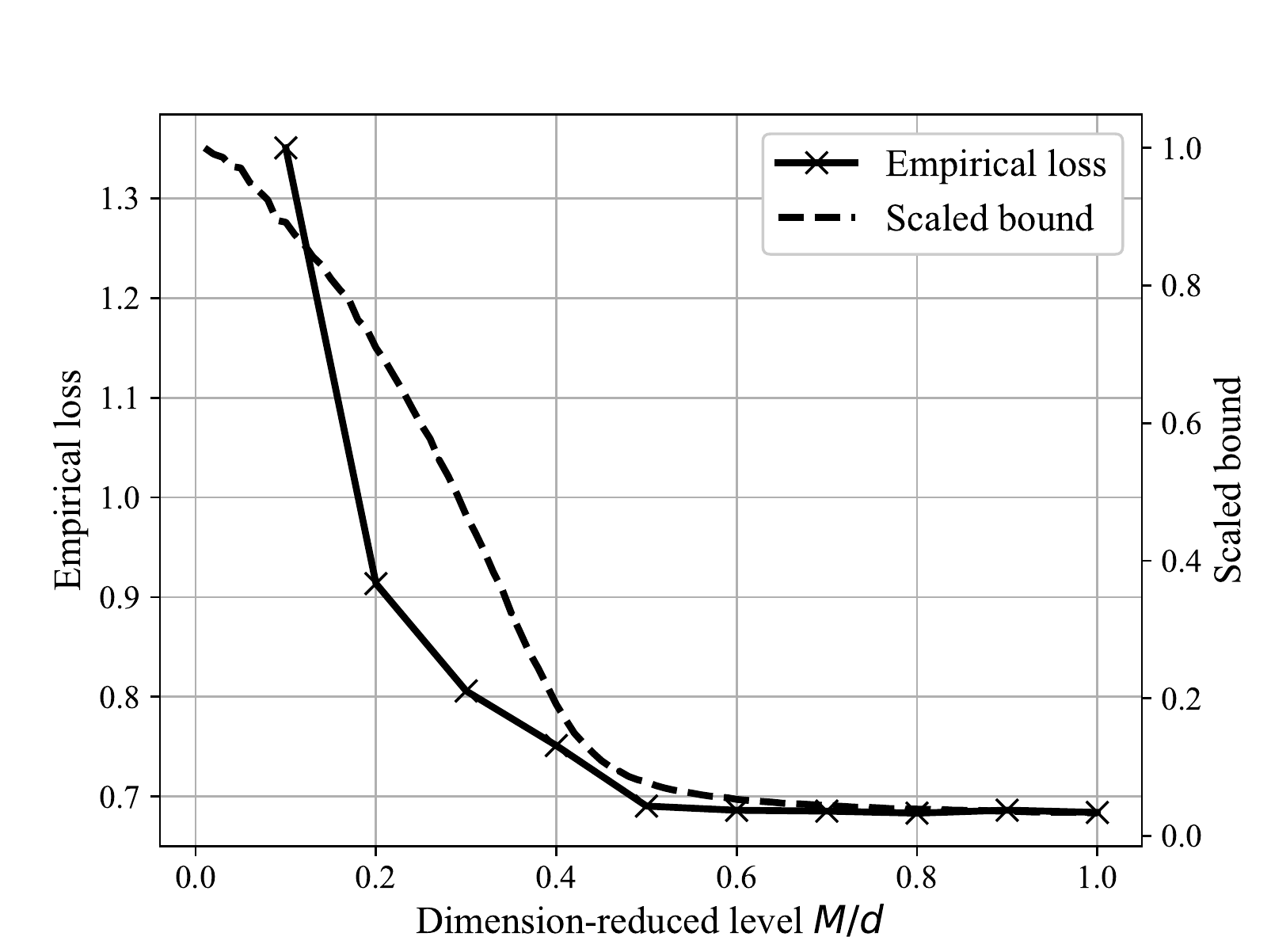}
    \vspace{-0.1in}
    \caption{Empirical loss performance of AirFL-Clip-Comp versus $M/d$ under $k/d=0.01$ and $\eta^{(t)}=120/(400+t)$, along with a scaled version of the theoretical bound in  \eqref{eq:bound} versus $M/d$. To evaluate  \eqref{eq:bound}, we set $G=100$ and $L=100$, {\color{black} and evaluate $v^{(t)}$ via SE of OAMP.}}
    \label{fig:loss_vs_M_d}
    \vspace{-0.1in}
\end{figure}


\section{Conclusions}
Previous work on AirFL proved the optimality of power control via norm clipping -- referred to here as AirFL-Clip -- for AWGN channels with instantaneous per-device power constraints under smooth loss functions satisfying PL condition. This paper has offered two contributions to the study of AirFL-Clip. First, we have introduced a novel convergence error bound  that applies to general smooth and non-convex learning objectives, and meanwhile supports offline evaluation. Second, we have extended AirFL-Clip to include Top-$k$ sparsification and linear compression (AirFL-Clip-Comp), for which we have also presented a convergence error bound for general smooth and non-convex learning objectives. 

Future work may focus on analyzing the impact of fading, and on investigating the interplay of clipping with Bayesian federated learning protocols.


\appendix[Sketch of the Proof of Theorem \ref{thm:convergence}]

To prove Theorem \ref{thm:convergence}, we need to introduce the following  lemmas.
\begin{lemma}[Top-$k$ Contraction] \label{lemma:Top-K}
For a parameter $0<k<d$, $\text{Top}_k: \mathbb{R}^{d\times 1} \to \mathbb{R}^{d\times 1}$  satisfies the contraction property 
\begin{equation}
    \|\mv{x}-\text{Top}_k(\mv{x})\|^2 \le \left( 1-\lambda \right)\|\mv{x}\|^2,
\end{equation}
where we recall the definition $\lambda=k/d$, and equality is attained when all entries of vector $\mv x$ are equal.
\end{lemma}
\begin{IEEEproof}
   This can be proved by contradiction.
\end{IEEEproof}

\begin{lemma}[Memory Contraction {\cite[Lemma 4]{basu2020qsparse}}] \label{lemma:memory_bound}
    For $a>\frac{4Q}{\lambda},\eta^{(t)}=\frac{\xi}{a+t}$, and $t\in \mathbb{Z}^+$, there exists a constant $C\ge \frac{4a\lambda(1-\lambda^2)}{a\lambda-4Q}$ such that
\begin{equation}
    \|\mv{m}_{i}^{(t)}\|^2 \le 4 \frac{(\eta^{(t)})^2}{\lambda^2}CQ^2G^2.
\end{equation}
\end{lemma}

Based on Lemma \ref{lemma:Top-K}, we have $\|\mv{g}_{i}^{(t)}\|^2 \le \left( \frac{8C}{\lambda^2}+2\right)(\eta^{(t)})^2Q^2G^2$ by the fact that $\|\mv{g}_{i}^{(t)}\|^2 \le \| \mv{m}_{i}^{(t)}+\mv \Delta_i^{(t)} \|^2 $ with the model difference bound $\|\mv \Delta_i^{(t)} \|^2 \le (\eta^{(t)})^2Q^2G^2$ and using $\|\mv a+\mv b\|^2 \le 2\|\mv a\|^2 + 2\|\mv b\|^2$.

Beginning with Assumption \ref{assumption:L-smoothness} and the global update rule in \eqref{eq:clip global updatas}, taking expectation conditioned on $\mv \theta^{(t)}$, we separate the noise term \(\mv n_e^{(t)}\) using Assumption \ref{assumption:Estimation Error}, yielding
\begin{multline} \label{eq:L_inequality_global_update}
     \mathbb{E}\left[f\left(\mv \theta^{(t+1)}\right)\right]  \leq f\left(\mv \theta^{(t)}\right)+\frac{L}{2} \mathbb{E}\left[\left\|\frac{1}{R} \sum_{i = 1}^{R} \mv \delta_i^{(t)}\right\|^{2}\right]+\\
    \left\langle\nabla f\left(\mv \theta^{(t)}\right), \mathbb{E}\left[\frac{1}{R} \sum_{i = 1}^{R} \mv \delta_i^{(t)}\right]\right\rangle +\frac{L d}{2R^2} (\eta^{(t)})^2  v^{(t)},
\end{multline}
where $\mv \delta_i^{(t)} = -\eta^{(t)} \tilde{\mv x}_i^{(t)}=-\mv g_i^{(t)} \alpha_i$ with clipping factor $\alpha_{i}^{(t)}:=\min\left\{1,\sqrt{P}/\left(\left\|\mv{g}_i^{(t)}/\eta^{(t)}\right\|\right)\right\}$ and $\Gamma\le \alpha_i^{(t)}\le 1$, which can be derived by Assumption \ref{assumption:Bounded Instantaneous Norm} and $\min\{a, b\} = ab/ \max\{a, b\} \ge ab/(a + b)$. Define $\bar{\alpha}^{(t)}=\frac{1}{R}\sum_{i=1}^R \mathbb{E}[\alpha_{i}^{(t)}]$.

For the inner product term, using $\langle \mv a,\mv b\rangle=-\frac{1}{2}\|\mv a\|^{2}-\frac{1}{2}\|\mv b\|^{2}+\frac{1}{2}\|\mv a-\mv b\|^{2}$, $L$-smoothness, Lemma \ref{lemma:memory_bound}, the update rule of memory, and decoupling $\mv{g}_{i}^{(t)}$ and $\bar{\alpha}_i$ by using Assumption \ref{assumption:Bounded Instantaneous Norm}, leveraging results in \cite[Lemma 3]{reddi2021adaptive} (according to Assumption \ref{assumption:Bounded Instantaneous Norm} and \ref{assumption:Bounded Data Heterogeneity}) and the upper bounds on $\|\mv m_{i}^{(t)}\|$ and $\|\mv g_i^{(t)}\|$, we have the upper bound of the inner product term. Specifically, with the aid of Jensen's inequality and Cauchy-Schwartz inequality and with the definition $\eta^{(t)} = \frac{\xi}{a+t} \leq \frac{1}{\sqrt{120} Q L}$, i.e., $a\ge \sqrt{120}\xi QL$, we have
\begin{multline} \label{eq:inner_product_bound}
    \left\langle\nabla f\left(\mv \theta^{(t)}\right), \mathbb{E}\left[\frac{1}{R} \sum_{i = 1}^{R} \mv{\delta}_{i}^{(t)}\right]\right\rangle\le 2\eta^{(t)} Q G^2\sqrt{ \frac{8C}{\lambda^2}+2}+\\
     \frac{\eta^{(t)} \bar{\alpha}^{(t)}}{2} \Bigg(10L^{2} Q^{2} (\eta^{(t)})^2\left(\sigma_{l}^{2}+6 Q \sigma_{g}^{2}\right) + \frac{32QC}{\lambda^2}G^2\Bigg)+\\
    \frac{-\eta^{(t)} \bar{\alpha}^{(t)} Q}{4}\left\|\nabla f\left(\mv \theta^{(t)}\right)\right\|^{2}.
\end{multline}

For the quadratic of the norm term, we use $\|\mv a + \mv b + \mv c\|^2 \le 3\|\mv a\|^2 + 3\|\mv b\|^2 + 3\|\mv c\|^2$, $0<\alpha_i^{(t)}<1$, and Jensen's inequality, to obtain
\begin{equation} \label{eq:quadratic_of_the_norm_bound}
    \begin{aligned}
        \mathbb{E}\left[\left\|\frac{1}{R}\sum_{i=1}^{R} \mv{\delta}_{i}^{(t)}\right\|^{2}\right] \le 9\left( \frac{8C}{\lambda^2}+2\right)(\eta^{(t)})^2Q^2G^2.
    \end{aligned}
\end{equation}

Substituting \eqref{eq:inner_product_bound} and \eqref{eq:quadratic_of_the_norm_bound} to \eqref{eq:L_inequality_global_update}, taking summation over $t$ from $0$ to $T-1$, rearranging, using $\sum_{t=0}^{T-1}\eta_t  \mathbb{E}\left[\Gamma \left\|\nabla f\left(\mv \theta^{(t)}\right)\right\|^{2}\right] \le \sum_{t=0}^{T-1}\eta_t  \mathbb{E}\left[\bar{\alpha}^{(t)}\left\|\nabla f\left(\mv \theta^{(t)}\right)\right\|^{2}\right]$, dividing both sides by $\Gamma P_T Q / 4$, and taking an expectation over all random variables, where $P_T=\sum_{t=0}^{T-1} \eta^{(t)} \geq \xi \ln \left(\frac{T+a-1}{a}\right)$, we obtain \eqref{eq:bound} with the aid of $\sum_{t=0}^{T-1}(\eta^{(t)})^2 \le \frac{\xi^2}{a-1} $ and $\sum_{t=0}^{T-1}(\eta^{(t)})^3 \le \frac{\xi^3}{2(a-1)^2}$.

\addtolength{\topmargin}{-0.02in}
\bibliographystyle{IEEEtran}
\bibliography{DL_ref}

\end{document}